\documentclass[12pt,epsf,epsfig]{article}
\usepackage{epsfig}

\begin{document}

\title{Constraints on Weakly Mixed Sterile Neutrinos in the Light of SNO Salt Phase
and 766.3 Ty KamLAND Data}
\author{S. Dev\thanks{%
E--mail : dev5703@yahoo.com} and Sanjeev Kumar\thanks{%
E--mail : sanjeev3kumar@yahoo.co.in } \\
\emph{Department of Physics, Himachal Pradesh University,} \\
\emph{Shimla, India-171005.}}
\maketitle

\begin{abstract}
The possibility of flavor transitions into sterile neutrinos (accompanying
the dominant LMA transitions) in the solar boron neutrino flux has been
examined in a scenario proposed by Hollanda and Smirnov to overcome some
generic problems of the pure LMA scenario. It is found that the most recent
SNO salt phase solar neutrino data and the KamLAND 766.3 Ty spectral data,
allow for a significant sterile presence in the solar boron neutrino flux
reaching the earth.
\end{abstract}


\section{Introduction}

The SNO \cite{1,2} and KamLAND \cite{3} experiments have played a
crucial role in resolving the longstanding solar neutrino problem
in terms of large mixing angle (LMA) MSW oscillations and are
expected to play an important role in the refinement of the LMA
solution which is undergoing a deeper scrutiny. Does the LMA
solution explain all the solar neutrino data satisfactorily? There
are, at least, two generic predictions of LMA indicating new
physics beyond LMA. One of these is the prediction of the high
argon production rate for Homestake experiment which is about
2$\sigma $ above the observed rate. Another generic prediction of
the LMA scenario is the `spectral upturn' at low energies. Within
the LMA parameter space, the survival probability should increase
with decrease in energy and for the best fit parameters, the
upturn could be as large as 10-15\% between 8 MeV and 5 MeV
\cite{4}. In fact, the spectral upturn at low energies is expected
to increase further with the KamLAND 766.3 Ty spectral data
\cite{5} favoring a larger value of $\Delta m^{2}$ \cite{6}.
However, neither SuperKamiokande nor SNO has reported any
statistically significant `rise-up' in the observed neutrino
survival probability. Both these predictions of the LMA solution
can, only, be tested in the forthcoming phase of high precision
measurements \cite{7} in the solar neutrino experiments and are
crucial for the final confirmation of the LMA solution.

Another unresolved issue is whether the solar neutrinos oscillate into the
sterile component. The main motivation for postulating the existence of the
sterile neutrino species comes from the LSND experiment which reported a
significant $\overline{\nu }_{\mu }\rightarrow \overline{\nu }_{e}$
oscillation probability \cite{8} which requires a new mass scale. Since, the
Z-decay width constrains the number of weakly interacting light neutrino
species to be very close to three \cite{9}, one is forced to postulate a
sterile neutrino. While the purely sterile oscillation solution is excluded
at 7.6$\sigma $ \cite{10}, the solar electron neutrinos could still
oscillate into both active and sterile neutrinos, a scenario which is,
largely, unconstrained at present. In fact, a combined analysis \cite{11} of
solar and atmospheric neutrino data has shown that the active-sterile
admixture can take any value between zero and one. While the SNO charged
current data excluded the maximal mixing to sterile neutrinos at 5.4$\sigma $
\cite{2}, arbitrary active-sterile admixtures were not considered.
Consequently, a significant sterile fraction in the solar neutrino flux
reaching the earth is, still, possible. The discovery of the sterile
neutrinos would be of great importance for particle physics, even though, it
is, still, not clear how these hypothetical `exotic' degrees of freedom
would fit into elementary particle theory.

The possibility of subdominant transitions into sterile neutrino states
accompanying the dominant LMA flavor transitions has been examined earlier
\cite{12} and upper bounds on the sterile neutrino fraction in the
non-electronic boron neutrino flux have been derived. However, the
subdominant transitions into sterile states have neither been confirmed nor
ruled out at a statistically significant level. In the present work, the
possibility of flavor transitions into sterile component in the solar boron
neutrino flux has been examined in a model presented by Hollanda and Smirnov
\cite{4} to lower the abnormally high argon production rate in the Homestake
experiment and, also, to lower the `spectral upturn' in the low energy boron
neutrino spectrum predicted in the pure LMA scenario.

\section{Weakly Mixed Sterile Neutrinos}

We introduce a sterile neutrino $\nu _{s}$ which mixes weakly with the
active flavors $\nu _{e}$ and $\nu _{\mu }$ to form the mass eigenstates $%
\nu _{0}$, $\nu _{1}$ and $\nu _{2}$ given by
\begin{eqnarray}
\nu _{0} &=&\left( \cos \alpha \right) \nu _{s}+\sin \alpha \{\left( \cos
\theta \right) \nu _{e}-\left( \sin \theta \right) \nu _{\mu }\},  \nonumber
\\
\nu _{1} &=&\left( \cos \alpha \right) \{\left( \cos \theta \right) \nu
_{e}-\left( \sin \theta \right) \nu _{\mu }\}-\left( \sin \alpha \right) \nu
_{s},  \nonumber \\
\nu _{2} &=&\left( \sin \theta \right) \nu _{e}+\left( \cos \theta \right)
\nu _{\mu },
\end{eqnarray}
with masses $m_{0}$, $m_{1}$ and $m_{2}$, respectively. It is assumed that $%
\sin ^{2}\alpha \ll 1$ (weak mixing) so that $\nu _{s}$ is mainly present in
the mass eigenstate $\nu _{0}$ only. Following Hollanda and Smirnov \cite{4}%
, we assume the mass hierarchy $m_{1}<\ m_{0}<m_{2}$, and define the
following mass squared differences:
\begin{eqnarray}
\Delta m_{01}^{2} &=&m_{0}^{2}-m_{1}^{2},  \nonumber \\
\Delta m_{12}^{2} &=&m_{2}^{2}-m_{1}^{2}.
\end{eqnarray}
The energy eigenlevels for the above neutrino system ($\nu _{0}$, $\nu _{1}$
and $\nu _{2}$) are denoted by $\lambda _{0}$, $\lambda _{1}$ and $\lambda
_{2}$, respectively. For the mass hierarchy assumed above, the level $%
\lambda _{0}$ crosses the level $\lambda _{1}$ only and $\lambda _{2}$ is,
approximately, the same as it would be in the pure LMA two flavor scenario
in the absence of any sterile mixing. Neglecting the small admixture of $\nu
_{e}$ in $\nu _{0}$, one obtains
\begin{eqnarray}
P_{ee} &=&P_{LMA}-P_{es}\cos ^{2}\theta , \\
P_{e\mu } &=&1-P_{LMA}-P_{es}\sin ^{2}\theta , \\
P_{es} &=&\cos ^{2}\theta _{m}(\sin ^{2}\alpha _{m}+P_{c}\cos 2\alpha _{m}),
\end{eqnarray}
where, $P_{LMA}$, given by
\begin{equation}
P_{LMA}=\frac{1}{2}+\frac{1}{2}\cos 2\theta \cos 2\theta _{m},
\end{equation}
is the survival probability for electron neutrinos in the pure LMA scenario.
$P_{c}$ is the crossing probability at the point where $\lambda _{0}$ and $%
\lambda _{1}$ cross while $\theta _{m}$ and $\alpha _{m}$ are the mixing
angles in the matter. The symbols $P_{ee}$, $P_{e\mu }$ and $P_{es}$ have
their usual meaning. Hollanda and Smirnov \cite{4} have shown that the
introduction of sterile admixture leads to a decrease in the `rise-up' in
the boron neutrino spectrum at lower energies and, also, reduces the argon
production rate at Homestake to a phenomenological acceptable level for
\begin{eqnarray}
\Delta m_{01}^{2} &\sim &\left( 2-20\right) \times 10^{-5}eV^{2},  \nonumber
\\
\sin ^{2}2\alpha &\sim &\left( 10^{-5}-10^{-3}\right) .
\end{eqnarray}
Apart from some rather `exotic' scenarios proposed in literature \cite{13},
it happens to be the simplest and the most plausible scenario to overcome
the generic problems of the pure LMA solution mentioned earlier. Therefore,
it is important to constrain the sterile component in this scenario
(referred to as the (LMA+sterile) scenario, henceforth) in the light of the
SNO solar neutrino data.

In this (LMA+ sterile) scenario,
\begin{equation}
P_{ee}=\frac{\phi _{CC}^{SNO}}{\phi _{B}},
\end{equation}
\begin{equation}
P_{e\mu }=\frac{\phi _{NC}^{SNO}-\phi _{CC}^{SNO}}{\phi _{B}},
\end{equation}
\begin{equation}
P_{es}=1-\frac{\phi _{NC}^{SNO}}{\phi _{B}},
\end{equation}
where $\phi _{CC}$ and $\phi _{NC}$ are the fluxes measured at SNO through
CC and NC reactions, respectively, and $\phi _{B}$ is the total boron
neutrino flux.

From equations (8-10), one obtains
\begin{equation}
P_{e\mu }=\frac{1-x}{x}P_{ee}
\end{equation}
and
\begin{equation}
P_{es}=1-\frac{P_{ee}}{x}
\end{equation}
where
\begin{equation}
x=\frac{\phi _{CC}^{SNO}}{\phi _{NC}^{SNO}}.
\end{equation}
The ratio of nonelectronic active neutrino flux to total nonelectronic
(active+sterile) neutrino flux, denoted by $\sin ^{2}\varphi $, is given by
\begin{equation}
sin^{2}\varphi =\frac{\phi _{NC}^{SNO}-\phi _{CC}^{SNO}}{\phi _{B}-\phi
_{CC}^{SNO}}.
\end{equation}
In the LMA scenario
\begin{equation}
P_{LMA}\left( \nu _{e}\rightarrow \nu _{e}\right) =P_{LMA}=x,
\end{equation}
where $x$ is given by equation (13). However, in the (LMA +sterile)
scenario, $\phi _{B}$ and $\phi _{NC}^{SNO}$ are not equal and one has to
use the relation $(8)$%
\[
P_{ee}=\frac{\phi _{CC}^{SNO}}{\phi _{B}},
\]
instead, where $\phi _{B}$ is, now, an independent quantity which can not be
determined from the SNO CC and NC fluxes. One can use the boron neutrino
flux given by the standard solar model (SSM) for $\phi _{B}$ to calculate $%
sin^{2}\varphi $ and $P_{ee}$. However, because of large errors in the SSM
boron neutrino flux ($\phi _{SSM}$), only a lower bound on $sin^{2}\varphi $
can be obtained while the upper bound becomes larger than unity \cite{12}.

Without assuming $P_{ee}$, one cannot calculate $sin^{2}\varphi $ and $\phi
_{B}$, uniquely and only a family of solutions corresponding to different
values of $P_{ee}$ is obtained. Equations (8) and (14) can be rewritten as a
set of coupled equations as follows:
\begin{equation}
\sin ^{2}\varphi =\frac{1-x}{x}\frac{P_{ee}}{1-P_{ee}},
\end{equation}
and
\begin{equation}
\phi _{B}=\frac{\phi _{CC}^{SNO}}{P_{ee}}.
\end{equation}
This degeneracy is well known as the ($f_{B}-sin^{2}\varphi $) degeneracy in
the literature \cite{12}. The value of $\phi _{B}$ is, usually, given in the
units of the central value of $\phi _{SSM}$, so that $f_{B}=\phi _{B}/\phi
_{SSM}$).

To gain further insight into this degeneracy, we rewrite equations (11,12)
as follows
\begin{equation}
(1-x)P_{ee}-xP_{e\mu }=0
\end{equation}
and
\begin{equation}
P_{ee}+xP_{es}=x.
\end{equation}
Since, we have only two equations relating three unknowns viz. $P_{ee}$, $%
P_{e\mu }$ and $P_{es}$, a unique solution is not possible and one obtains a
family of solutions, instead, corresponding to different values of $P_{ee}.$
Balentekin et al. \cite{13} identify $P_{ee}$ with $P_{LMA}$ (electron
neutrino survival probability in the pure LMA scenario in the absence of any
sterile transitions) and use equations (8), (14) and (15) to constrain the
sterile component. However, equation (15) cannot be used to derive
meaningful constraints on the sterile component since one obtains $\sin
^{2}\varphi =1\ $on substitution of equation (15) in equation (16). To
overcome this problem, we use equation (3) alongwith equations (18,19) to
constrain the sterile component. We collect all these equations below:

\begin{eqnarray}
P_{ee}+P_{es}\cos ^{2}\theta &=&P_{LMA},  \nonumber \\
\left( 1-x\right) P_{ee}-xP_{e\mu } &=&0,  \nonumber \\
P_{ee}+xP_{es} &=&x.
\end{eqnarray}
This set of coupled equations has the following simultaneous solution
\begin{eqnarray}
P_{ee} &=&x\frac{\cos ^{2}\theta -P_{LMA}}{\cos ^{2}\theta -x}, \\
P_{e\mu } &=&\left( 1-x\right) \frac{\cos ^{2}\theta -P_{LMA}}{\cos
^{2}\theta -x}, \\
P_{es} &=&\frac{P_{LMA}-x}{\cos ^{2}\theta -x}.
\end{eqnarray}

\section{Results and Discussion}

In order to examine the possibility of transitions into sterile neutrinos,
we plot the $1\sigma $ allowed upper and lower values of $P_{ee}=\frac{\phi
_{CC}^{SNO}}{\phi _{B}}$ [equation (8)] allowed by the salt phase SNO data
\cite{2} and BP04 \cite{15} in Figure 1 which, also, depicts $P_{ee}$ as a
function of $P_{es}$ as given by equation (12). It is clear from Figure 1
that significant transitions into sterile neutrinos are allowed by the SNO
salt phase data and, in fact, the $1\sigma $ upper bound on $P_{es}$ could
be as large as 0.4. More precise bounds on the sterile fraction in the boron
neutrino flux can, only, be obtained with more precise measurements of CC
and NC rates at SNO in the future. It may be pertinent to mention here that
the boron neutrino flux estimates in the SSM have, rather, large
uncertainties. Consequently, considerable improvements in the boron neutrino
flux estimates in the SSM are required for obtaining meaningful constraints
on the possible sterile neutrino fraction in the boron neutrino flux.

One can obtain the electron neutrino survival probability $P_{ee}$, the
transition probability into muon neutrinos $P_{e\mu }$ and the transition
probability into the sterile neutrinos $P_{es}$ from equations (21-23) by
substituting the flux-averaged value of $P_{LMA}$ (calculated for the values
of $\Delta m^{2}$ and $\theta $ and $1\sigma $ errors therein taken from
\cite{16}) and the value of $x$ reported by SNO \cite{2}. It is important to
realize here that the SNO CC flux is the actual electron neutrino flux $\phi
_{\nu _{e}}$\ if the boron energy spectrum is assumed to be undistorted.
Since, the SuperKamiokande \cite{17} (with a better precision) has not
reported any, statistically significant, spectral distortions in the boron
neutrino spectrum, we assume an undistorted boron neutrino spectrum and
identify $P_{LMA}$ with $x$ at the high energy end. However, since the LMA
scenario predicts a significant spectral upturn at lower energies \cite{6}, $%
P_{LMA}$ is expected to be, significantly, different from $x$ at lower
energies. For the flux averaged value of $P_{LMA}$, $P_{ee}=\allowbreak
0.239_{-0.055}^{+0.063}$, $P_{e\mu }=\allowbreak 0.543_{-0.048}^{+0.042}$
and $P_{es}=0.\,218_{-0.105}^{+0.103}$, which is non-zero at about $%
2.1\sigma $. The probabilities $P_{ee}$, $P_{e\mu }$ and $P_{es}$ have been
plotted in Figure 2 as functions of $x$ where the $1\sigma $ upper and lower
bounds have, also, been shown. It is clear from Figure 2 that the $1\sigma $
values of $P_{ee}$ and $P_{es}$ are of comparable magnitude. In fact, for
the smaller values of x, the transition probability into the sterile
neutrinos can, even, be larger than the electron neutrino survival
probability. It is, also, clear from Figure 2, that the errors in the value
of x are the most significant sources of error in the values of the
probabilities and once the value of x is settled by the experiments, the
errors in the probabilities will become much smaller. For example, for a
most conservative choice of $x=0.341$, the transition probability into the
sterile neutrinos is found to be $0.146_{-0.033}^{+0.040}$ which is non-zero
at $4.4\sigma C.L.$ If the value of x is settled below this value ( i.e. $%
x\leq 0.341$, as is most likely) by the experiments, the transition
probability into the sterile neutrinos as well as the corresponding
confidence level will be larger than the above values (at $x=0.341$). Thus,
the main source of error in the transition probability into the sterile
neutrinos being the uncertainty in the value of x, a more precise
determination of the value of x (below the value 0.341) will give a non-zero
value of $P_{es}$ at more than 4.4 standard deviations.

From equations (16) and (17), one obtains
\begin{equation}
sin^{2}\varphi =\frac{\left( 1-x\right) \left( \cos ^{2}\theta
-P_{LMA}\right) }{\cos ^{2}\theta -x\left( 1+\cos ^{2}\theta -P_{LMA}\right)
},
\end{equation}
\begin{equation}
f_{B}=R_{NC}\frac{\cos ^{2}\theta -x}{\cos ^{2}\theta -P_{LMA}},
\end{equation}
where $f_{B}$ and $R_{NC}$ are the boron neutrino flux and SNO NC flux
normalized to the central SSM boron neutrino flux, respectively.\ The ($%
f_{B}-\sin ^{2}\varphi $)\ degeneracy is, thus, lifted by the use of
equation(3).

In Figure 3, we plot $\sin ^{2}\varphi $ (equation (24)) as a function of x.
The active neutrino fraction, $\sin ^{2}\varphi $, increases with the
increase in the SNO CC flux and decrease in the SNO NC flux. Thus, the
sterile fraction in the active solar boron neutrino flux, $\cos ^{2}\varphi $%
, will be larger if the forthcoming measurements at SNO favor smaller values
of CC flux and larger values of NC flux. For $x=0.341$, $\sin ^{2}\varphi
=0\allowbreak .\,713_{-0.107}^{+0.125}$ and $f_{B}=\allowbreak \allowbreak
1.\,14_{-0.23}^{+0.29}$. Thus, the sterile fraction in the boron neutrino
flux is non-zero at about $2.3\sigma C.L.$ indicating oscillation of boron
neutrinos into sterile states in the current SNO data. The behavior of $\sin
^{2}\varphi $ with x in the present work is different from that reported by
Balentekin et al. \cite{14} by identifying $P_{ee}$ with $P_{LMA}$ and
substituting the numerical value of $P_{LMA}$ obtained in the pure $LMA$
scenario. However, as discussed earlier, such an approach can not be used to
derive meaningful constraints on the sterile fraction.

The boron neutrino flux obtained from equation (25) has been plotted as a
function of x in Figure 4. Most of the 1$\sigma $\ region of the boron
neutrino flux lies above its central value in the SSM in contradiction with
the value of boron neutrino flux in the pure LMA scenario.

\section{Conclusions}

In conclusion, the prospects for constraining the sterile neutrino fraction
in the born neutrino flux reaching the earth have been examined in a
scenario discussed by Hollanda and Smirnov \cite{4} to overcome some generic
problems of the LMA scenario. The indications for the presence of sterile
component in the boron neutrino flux in the light of the latest SNO salt
phase data and the 766.3 Ty KamLAND data are found to be strong enough to be
taken seriously. A precise determination of the CC and NC fluxes at SNO
within the present $1\sigma $ range gives a transition probability into
sterile states which is non-zero at more than $4.4\sigma $. It is found that
the sterile component in the boron neutrino flux could be as large as the
electron neutrino flux for the present central values of SNO CC and NC
fluxes. If the future measurements at SNO yield smaller values of CC/NC flux
ratio, the sterile fraction will be, further, enhanced. Thus, there are
strong indications of a sterile presence in the boron neutrino flux reaching
the detectors on the earth but it will require a precise determination of CC
and NC fluxes to pass a final judgment.

\section{Acknowledgments}

One of the authors (SK) gratefully acknowledges the financial support
provided by the Council for Scientific and Industrial Research (CSIR),
Government of India. The other author (SD) gratefully acknowledges the
financial support provided by the Department of Atomic Energy, Government of
India vide Grant No. 2004/37/23/BRNS/399.

\begin{figure}
\vspace{0.10in} \centerline{\epsfysize=6in\epsffile{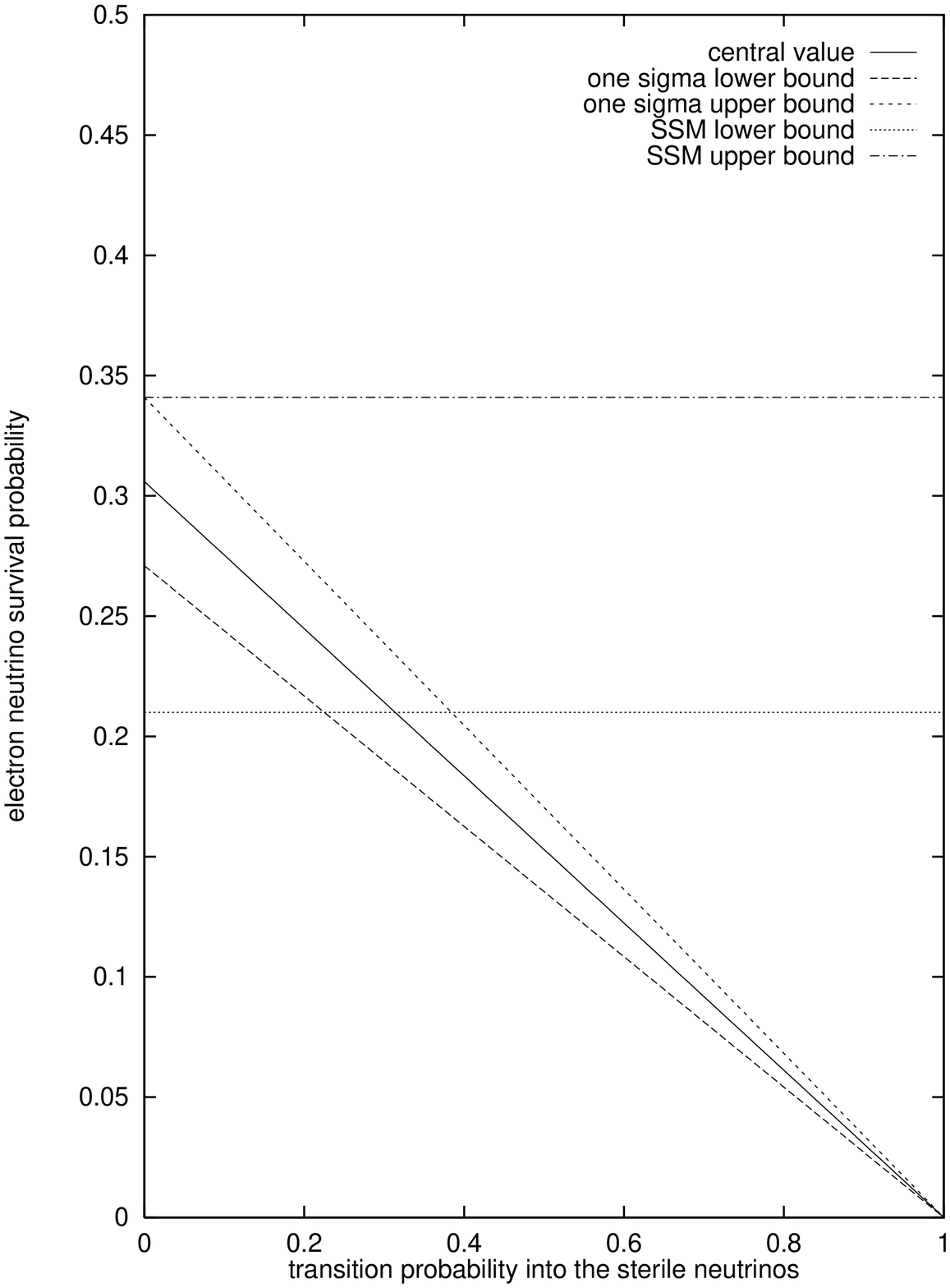}}
\vspace{0.08in} \caption{The $1\sigma $ allowed region in the
$P_{ee}-P_{es}$ space.} \label{fig1}
\end{figure}

\begin{figure}
\vspace{0.10in} \centerline{\epsfysize=6in\epsffile{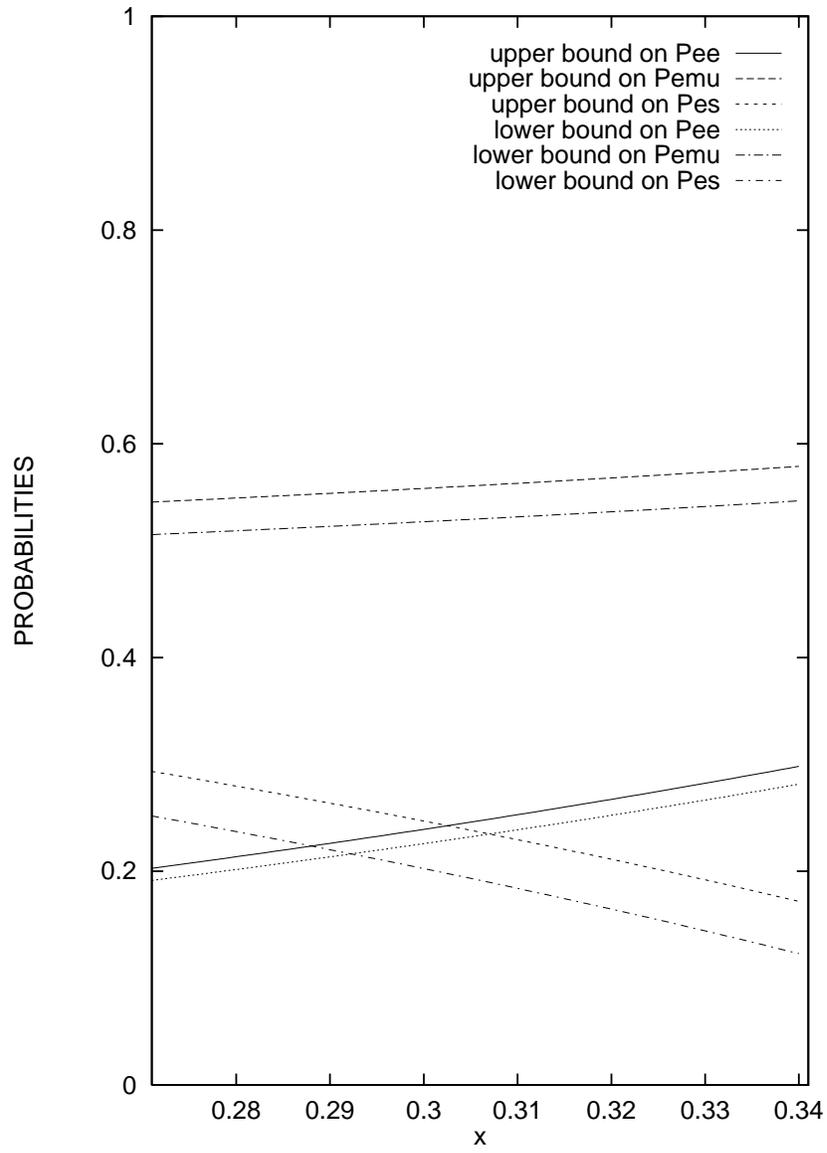}}
\vspace{0.08in} \caption{$P_{ee}$, $P_{e\mu }$ and $P_{es}$ as
functions of $x$.} \label{fig2}
\end{figure}

\begin{figure}
\vspace{0.10in} \centerline{\epsfysize=6in\epsffile{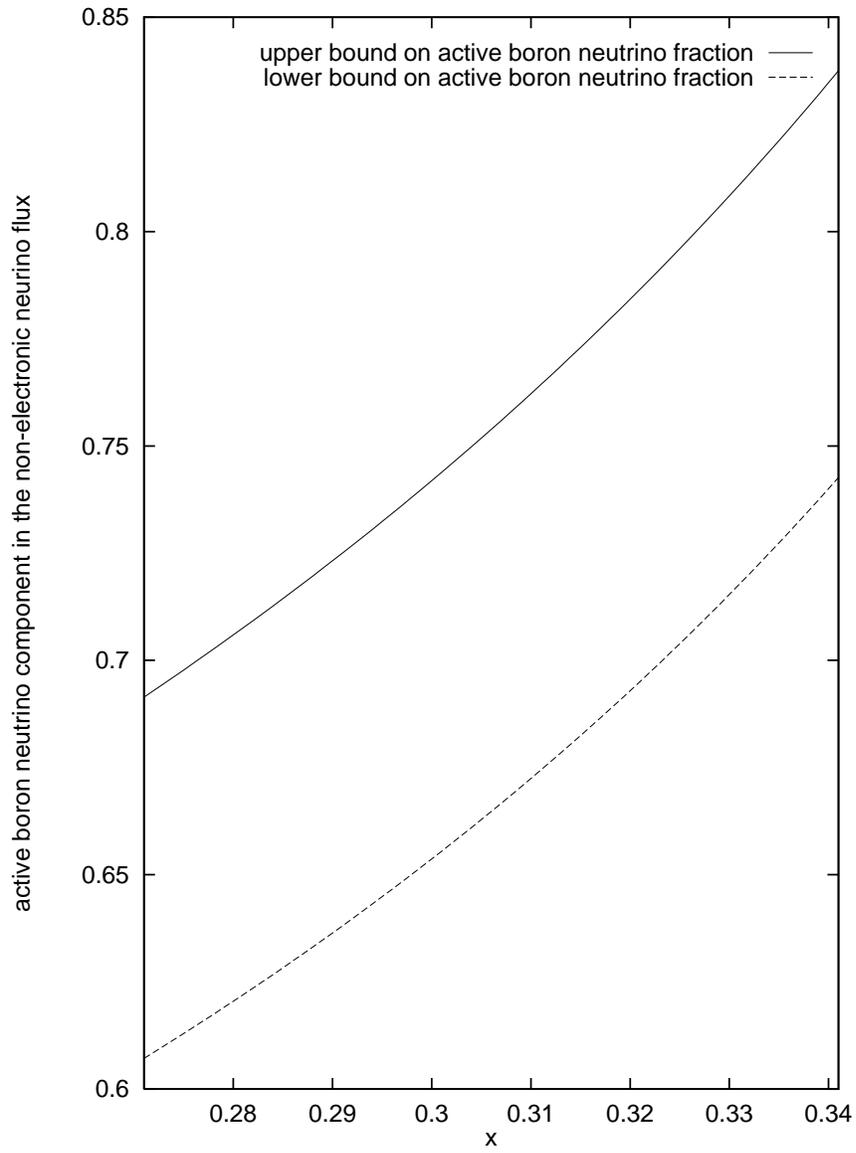}}
\vspace{0.08in} \caption{$\sin ^{2}\varphi $ as a function of $x$.
} \label{fig3}
\end{figure}

\begin{figure}
\vspace{0.10in} \centerline{\epsfysize=6in\epsffile{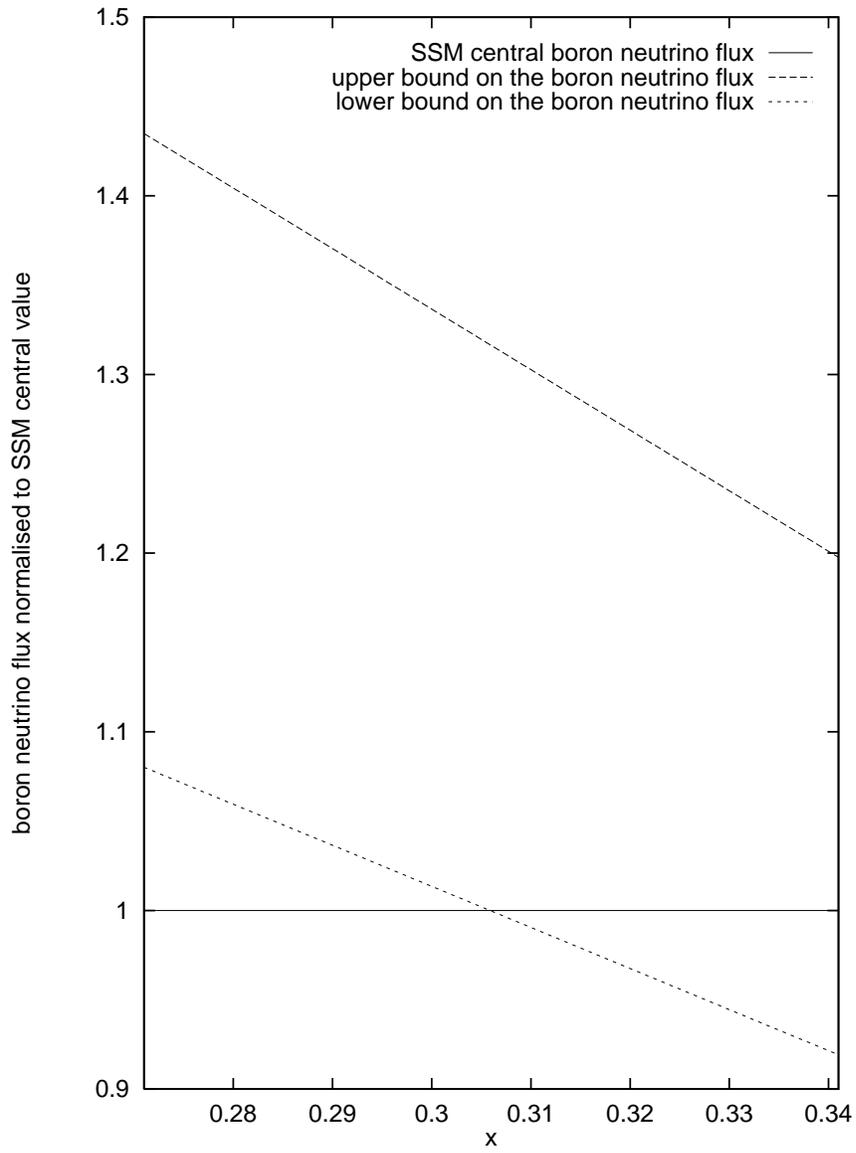}}
\vspace{0.08in} \caption{$f_{B}$ as a function of $x$.}
\label{fig4}
\end{figure}

\end{document}